\RequirePackage{lineno}
\documentclass[aps,showpacs,preprintnumbers,amsmath]{revtex4-2}
\usepackage{graphicx}
\usepackage{dcolumn}
\usepackage{epstopdf}
\usepackage{color}
\usepackage{tikz}
\usetikzlibrary{shapes.geometric}
\usepackage{bm}
\definecolor{dgreen}{cmyk}{1.,0.,1.,0.2}        
\definecolor{orange}{cmyk}{0.,0.353,1.,0.}    

\def\bea {\begin{eqnarray}}
\def\eea {\end{eqnarray}}

\def\be {\begin{equation}}
\def\ee {\end{equation}}

\begin{document}
\title{Existence of a second phase transition in A-A collisions at energies
  available at the CERN Large Hadron Collider: A physical interpretation}
\author{C. Pajares{$^1$}, R. P. Scharenberg{$^2$ } and B. K. Srivastava {$^2$}}
\medskip
\email {Contact: brijesh@purdue.edu}
\affiliation{$^1$Departamento de Fisica de Particulas, Universidale de Santiago de Compostela and Instituto Galego de Fisica de Atlas Enerxias(IGFAE), 15782 Santiago, de Compostela, Spain \\
$^2$Department of Physics and Astronomy, Purdue University, West Lafayette, IN-47907, USA}
\bigskip
\date{\today}

\begin{abstract}
The observed fluid behavior of the matter produced in heavy ion and $pp$ collisions implies that the strings 
stretched between the constituents of the projectile and target should have a hard core giving rise to a repulsion between them. It is shown that in these collisions there are two critical string densities, one corresponding to the well studied percolation of strings and another to the percolation of the hard core of the strings. These two critical densities are related to two critical temperatures, one T $\sim $  160 MeV would correspond to the restoration of chiral symmetry and partial deconfinement and a second transition T $\sim $ 220 MeV form the fluid behavior of QCD matter strongly interacting to a quasi free gas of quarks and gluons.
This argument is supported by the recently published results on the sudden increase in the degrees of freedom above the
deconfinement temperature in pp and A-A collisions \cite{aditya}.

\end{abstract}
\maketitle
\section{Introduction}
It is commonly accepted that Lattice Quantum Chromodynamics (LQCD) results \cite{aoki} showing a smooth
crossover in the properties of the thermal interacting matter at the
temperature T $\sim$ 155 MeV, going from confined matter to deconfined matter. This temperature coincides with the corresponding to the restoration of chiral symmetry.

On the other hand, the experimental results from RHIC \cite{arsene, back, adams, adcox} and LHC \cite{muller} indicates that a strongly interacting fluid medium is formed with extremely low shear viscosity over entropy density ratio $\eta/s$ in line with the results from the Ads/CFT holographic dual \cite{ads}.

However, the physics of thermal QCD transition is not quite settled and there
are several questions which should be addressed. In particular, the current accepted scenario offers limited room for describing the change from a weakly interacting hadron resonance gas to strongly interacting near perfect fluid, as well as how transition takes place from this state to a free gas of quarks and gluons.
In the literature there are several approaches pointing out new phases at low baryon density. 

In this paper we explore the possible signatures of this new phase and how it could be accommodated in the phenomenological models, particularly in the color string percolation model (CSPM) which describes rightly most of the RHIC and LHC experimental data \cite{review15}. In the CSPM the cross over transition from
hadronic phase to quarks and gluons phase at temperature $T\sim$ 160 MeV
corresponds to the critical string percolation. At this point a cluster of
overlapping strings is formed crossing the transverse surface of the collision.

The strings as chromoelectric superconductor repel each other analogous to the
interaction between vortex lines in a superconductor (Meissmer effect) meaning
that the strings have a hard core whose size is determined by the London
penetration depth which depends on the string tension \cite{paolo}. The existence of repulsion between strings and subsequently the existence of a hard core in the strings is also required by the radial distribution of the color strings in the
transverse collision surface. In fact, the radial distribution of sources of a fluid can not be periodic as in the case of a solid neither flat as in the case of a gas but some minimum and maximum like a hard core \cite{ramirez}.

 The size of the hard core provide us with a new scale. The percolation of these hard core of the strings, crossing the transverse collision surface, corresponds to a second critical string density and consequently to a second critical temperature, marking the transition from a fluid of quarks and gluons strongly interacting to a gas of free quarks and gluons. 

 In our earlier work the equation of state was obtained as shown  in Fig. 1 \cite{aditya}. In this figure the energy density is obtained using the well known Bjorken
 formula. The experimental data  of the ALICE Collaboration  for the multiplicities $dN/dy$ and the $p_{t}$ distribution in the range $ 0.15 < p_{t} < 2$ GeV is used to compute the dependence on the energy and centrality. The temperature is obtained by analyzing the  $p_{t}$ spectrum using the CSPM methodology. A sudden rise of $\varepsilon/T^{4}$ is seen at temperature $T \sim $ 220 MeV \cite{aditya}.

The plan of the paper is as follows : we start summarizing briefly some theoretical proposals of a new phase transition, namely, the different behavior of Dirac spectrum  in three different regions, the center vortex evidence of a second temperature above the crossover temperature, the existence of a stringy fluid phase
transition based on chiral spin symmetry, and the difference dependence on the power of $N_{c}$ of the energy , entropy densities of a confined hadron gas,
a stringy fluid and a liberated gluon gas.

In the next sections we introduce the Color String Percolation Model (CSPM) and present a new phase transition from a fluid to a free  gas of quarks and gluons. Finally the conclusions are presented.


 \section{Theoretical Proposals}
\subsection{Behavior of the Dirac Spectrum}
A new thermal phase of QCD  has been proposed, the infrared phase based in the analysis of the behavior of the Dirac spectrum \cite{alex1, alex2, alex3, alex4, alex5, meng}.
In quenched QCD increasing the temperature the properties of a thermal state remain
similar to the zero temperature vacuum, until the scale of thermal agitation
becomes comparable to the scale of broken scale invariance, i.e. gluon
condensate. This is characterized by the crossover temperature $T_{a}$. Beyond this temperature the properties of the medium change rapidly toward the restoration of the scale invariance at a well defined infrared
temperature $T_{ir} > T_{a}$.
In the range $T_{ir} < T <  T_{uv}$ (IR phase), the gauge fields are scale invariant at distances larger than $\frac {1}{T}$. The infrared
invariance appears due to the interaction that is still strong at long distances. For $ T > T_{uv}$ (UV phase), the field fluctuations of the IR phase disappear and the notion of the IR scale invariance becomes trivial.

The infrared temperature $T_{ir}$ coincides with the critical temperature of
the Polyakov line first order transition in quenched QCD. $T_{ir}$ is $\sim$
200-250 MeV. The crossover temperature $T_{a}$, $\sim$ 155 MeV is simply a characteristic temperature of the onset of changes toward the IR scale. Chiral
symmetry remains spontaneously broken for the range of temperatures
$T_{a} < T < T_{ir}$. Only after crossing such anomalous phase does the system become both deconfined and chirally symmetric.
 The lattice studies of the Dirac spectrum show that as the temperature cross
$T_{ir}$ the Dirac spectral function (the number of eigenmodes $\lambda$ per unit of volume and spectral interval) changes from extremely flat in the infrared
($\lambda$ close to zero) to follow a $1/\lambda$ curve \cite{alex3,alex4}.
At T $\sim $ 200 MeV well above the established region, the anomalous phase
fully develop with a clear peak and a large depletion at intermediate scales. This behavior is also seen not only in quenched QCD but also in the presence of light dynamical quarks. 
In this scheme it is argued that this IR phase is not detected by the lattice studies due to insufficient large volumes \cite{alex4}.

\begin{figure}[!h]
\centering        
\includegraphics[width=0.55\textwidth,height=3.5in]{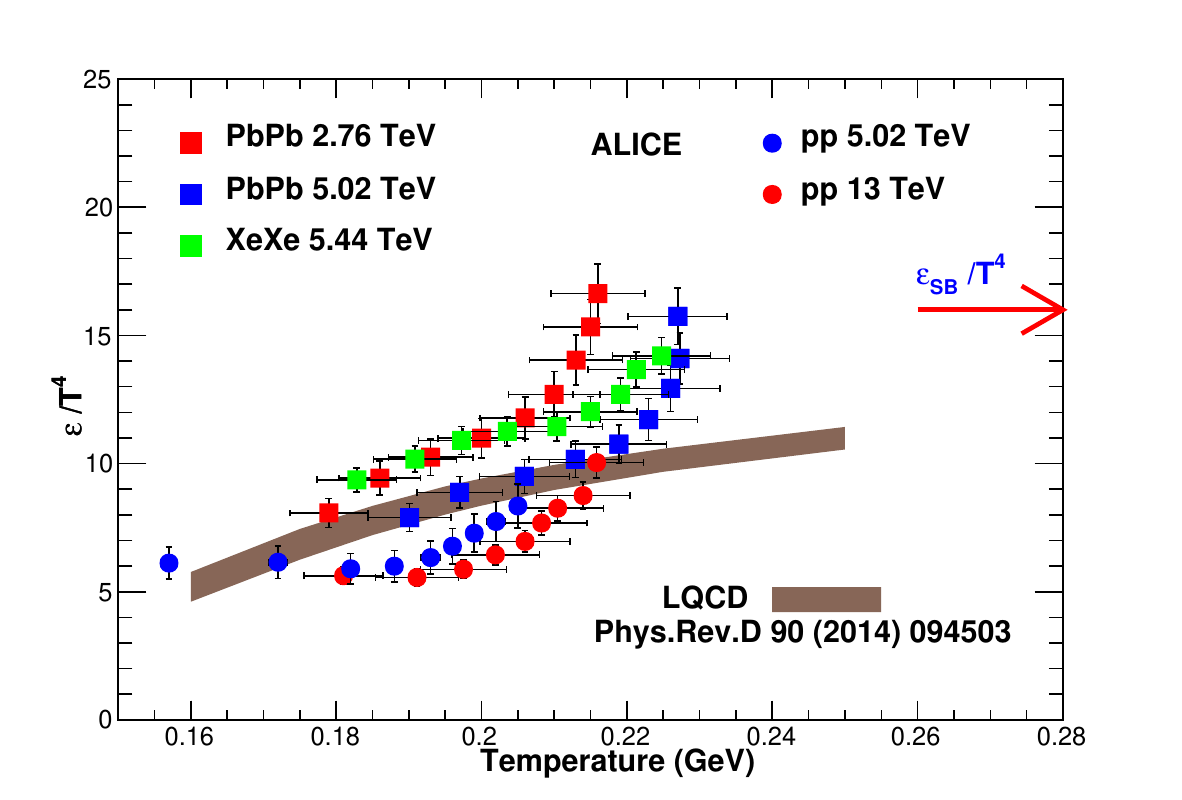}
 \vspace{-0.2cm}
\caption{ Dimensionless quantity $\bm \varepsilon/T^{4}$ as a function of temperature from CSPM using ALICE data \cite{aditya}. The LQCD calculations are from HotQCD Collaboration (brown band) \cite{lattice14}.}
\label{et4}
\end{figure}
\subsection{Center vortex evidence for a second finite-
  temperature QCD transition }
Center vortices are regions of the gauge fields that carry magnetic flux
 quantized
according to the center of SU(3). The center vortex structure is temporal and spatial slices of the lattice reveal that vortex percolation persists through the
chiral transition and ceases at a temperature that is approximate twice the chiral transition temperature $T_{c}$. This implies that confinement is retained through temperatures up to T around $ 2 T_{c}$, pointing toward a second transition corresponding to deconfinement. The loss of percolation is quantified by the vortex cluster event, providing a clear signal for the confinement transition \cite{vortex}. 

\subsection{Stringy Fluid}
In addition to the hadron gas and the quark gluon plasma phases, it was proposed the existence of a stringy fluid phase based on the chiral spin symmetry. At temperatures below the pseudocritical
temperature $T_{c} \sim $ 155 MeV the QCD matter is hadron gas with spontaneously broken chiral symmetry. Within the window $T_{c}$ -3$T_{c}$ the hot QCD is represented by stringy fluid with restored chiral and approximate chiral spin symmetries. Above $\sim 3T_{c}$ the chiral symmetry disappears and one observes a smooth transition to partonic degrees of freedom, i.e. to a quark-gluon plasma \cite{glozman1,glozman2,glozman3,philipsen}
\subsection{A new perspective on thermal transition in QCD}
Another proposal is the partial deconfinement 
which implies an intermediate phase in which color degree of freedom split into the confined and deconfined sectors \cite{watanabe,hanada,hanada2}. The partially deconfined phase is dual to the small black hole that lies between large black hole and the graviton gas. The transition to partial deconfinement would corresponds to the standard cross over. From partial to complete deconfinement would be accompanied by additional transition of the energy density. The three phases are characterized by the behavior of the function embedded in the phases of the Polyakov line.

\subsection{A New State of Matter between the Hadronic Phase and the Quark-Gluon Plasma? }
The energy density, pressure and entropy density in the stringy fluid scale as
$N_{c}$ which is in contrast to the independence of $N_{c}$ of a gas of color
confined hadrons and to the $N_{c}^{2}$ scaling of the liberated gluons. Therefore, as the temperature increases, matter should first go through a region where the
energy density is independent of $N_{c}$ and then transit to a partially deconfined phase with the energy density of the order $\mathcal{O}(N_{c})$, and should eventually reach a fully deconfined phase with gluons that lift the energy density of the order of $\mathcal{O} (N_{c}^{2})$. The proposed phase diagram would be like Fig. 2 where the intermediate phase is denoted by SQGB(Spaghetti of Quarks with Gluballs) \cite{sqgp1,sqgp2,larry}.

 Thermodynamics quantities have been evaluated using a string based model for the density of states. The obtained equation of state is shown in Fig.3 (Green square). The experimental energy points (CSPM) in Fig.3 are same as in Fig. 1. It is observed that experimental $\varepsilon/T^{4}$ is in agreement with lattice results \cite{lattice14} up to the temperature T $<$ 200 MeV and a departure from it above T$\sim$200 MeV. The CSPM result sees a clear increase of  $\varepsilon/T^{4}$ above T$\sim$200 MeV.

 There are more examples of non trivial behavior above $T_{c}$ , for instance, the spectral peak of the heavy quarkonium remains above $T_{c}$ \cite{asakwa,burnier,bala}. A partially confined/deconfined new phase has also been found in a model with the adjoint Polyakop loop \cite{myers}.

 \subsection{String Models}
 
 The above mentioned
proposals are different and therefore also their details as well as the values of temperature transitions and the nature and order of the transition. We will
focus in common general aspects of all of them as the increase of the degrees of freedom and correspondingly of the entropy and energy densities and the existence of two temperatures and the corresponding two scales.

Most of the models describing the RHIC and LHC data, like PYTHIA \cite{pythia}, EPOS \cite {epos}, AMPT \cite{ampt}, HIJING \cite{hijing}, and
Color String Percolation (CSPM) \cite{review15}, are based on the formation of strings stretched between projectiles and target. In the glasma picture of color glass
condensate, also are formed strings (color ropes) which are in correspondence with the clusters of strings of the CSPM \cite{cgc,cgc2}. Further discussions are focused on CSPM and interpretation  of the results as shown in Fig.~\ref{et4}.

\section{Color String Percolation Model}
Multiparticle production is currently described in terms of color strings stretched between the projectile and the target, which decay into new strings and subsequently hadronize to  produce the observed hadrons \cite{review15}. Color strings may be viewed as small areas in the transverse plane filled with color field created by colliding partons. In terms of gluon color field they can be considered as the color flux tubes stretched between the colliding partons. The mechanism of particle creation is the Schwinger $QED_{2}$ mechanism and is due to the color string breaking \cite{schw,wong}.
With growing energy and size of the colliding system, the number of strings grows, and they start to overlap, forming clusters, in the transverse plane very much similar to disks in two dimensional percolation theory \cite{inch,satzbook}. At a certain critical density, a macroscopic cluster appears that marks the percolation phase transition. This is the Color String percolation Model (CSPM) \cite{review15}.
The interaction between strings occurs when they overlap and the general result, due to the SU(3) random summation of charges, is a reduction in multiplicity and an increase in the string tension, hence an increase in the average transverse momentum squared, $\langle p_{T}^{2} \rangle$.
We assume that a cluster of ${\it n}$ strings that occupies an area of $S_{n}$ behaves as a single color source with a higher color field $\vec{Q_{n}}$ corresponding to the vectorial sum of the color charges of each individual string $\vec{Q_{1}}$. The resulting color field covers the area of the cluster. As $\vec{Q_{n}} = \sum_{1}^{n}\vec{Q_{1}}$, and the individual string colors may be oriented in an arbitrary manner respective to each other , the average $\vec{Q_{1i}}\vec{Q_{1j}}$ is zero, and $\vec{Q_{n}^2} = n \vec{Q_{1}^2} $.

Knowing the color charge $\vec{Q_{n}}$ one can obtain the multiplicity $\mu$ and the mean transverse momentum squared $\langle p_{t}^{2} \rangle$ of the particles produced by a cluster of $\it n $ strings \cite{pajares2}

\begin{equation}
\mu_{n} = \sqrt {\frac {n S_{n}}{S_{1}}}\mu_{0};\hspace{5mm}
\langle p_{t}^{2} \rangle = \sqrt {\frac {n S_{1}}{S_{n}}} {\langle p_{t}^{2} \rangle_{1}}
\label{mu}
\end{equation}
where $\mu_{0}$ and $\langle p_{t}^{2}\rangle_{1}$ are the mean multiplicity and $\langle p_{t}^{2} \rangle$ of particles produced from a single string with a transverse area $S_{1} = \pi r_{0}^2$. In the thermodynamic limit, one obtains an analytic expression \cite{pajares2,pajares1}
\begin{equation}
\langle \frac {n S_{1}}{S_{n}} \rangle = \frac {\xi}{1-e^{-\xi}}\equiv \frac {1}{F(\xi)^2};\hspace{5mm}
F(\xi) = \sqrt {\frac {1-e^{-\xi}}{\xi}}
\label{xi}
\end{equation}
where $F(\xi)$ is the color suppression factor. $\xi = \frac {N_{s} S_{1}}{S_{N}}$ is the percolation density parameter assumed to be finite when both the number of strings $N_{S}$ and total interaction area $S_{N}$ are large.  Eq.~(\ref{mu}) can be written as $\mu_{n}=F(\xi)\mu_{0}$ and 
$\langle p_{t}^{2}\rangle_{n} ={\langle p_{t}^{2} \rangle_{1}}/F(\xi)$.  
The critical cluster which spans $S_{N}$, appears for $\xi_{c} \ge$ 1.2 \cite{satzbook}.

The decay of Lund type string follows a Gaussian $p_{T}$ distribution, whose width is proportional to the string tension $x$.
\begin{equation}
 \frac{dn}{dp_{T}^{2}} \sim \exp (-\pi p_{T}^{2}/x^{2})
  \label{swing}
\end{equation}
The Gaussian fluctuations of the string tension give rise to a thermal
 distribution \cite{bialas}
  \begin{equation}
  \frac{dn}{dp_{T}^{2}} \sim \sqrt {\frac{2}{<x^{2}>}}\int_{0}^{\infty}dx \exp (-\frac{x^{2}}{2<x^{2}>}) \exp (-\pi \frac{p_{T}^{2}}{x^{2}})
\end{equation}

\begin{equation}
  \frac{dn}{dp_{t}^{2}} \sim \exp (-p_{t} \sqrt {\frac {2\pi}{\langle x^{2} \rangle}} ),
\end{equation}  
with $\langle x^{2} \rangle$ = $\pi \langle p_{t}^{2} \rangle_{1}/F(\xi)$.

The initial temperature is expressed as
\begin{equation}
T(\xi) =  {\sqrt {\frac {\langle p_{t}^{2}\rangle_{1}}{ 2 F(\xi)}}},
\label{temp}
\end{equation}
This temperature can be regarded as a local one associated to the transverse momentum distribution of particles produced in the fragmentation of a cluster of
strings. However, above the critical string percolation, the cluster is extended all over most of the transverse surface of the collision and thus, the local
 temperature can be considered as a global temperature.

 The color string percolation model describes rightly most of the experimental
 RHIC and LHC data, including rapidity and transverse momentum distributions, multiplicity distributions, long range correlations, ridge structure, $p_{t}$ correlations, and azimuthal distributions \cite{review15}. The mechanism to
 describe these distributions is loss of momentum of a parton produced in the
 fragmentation of a cluster. This is due to the interaction with the color
 field of the rest of strings of the collision surface in the way to get out the collision surface. This mechanism predicted the scaling law
 $v_{2} \propto p_{t}^{2/3}$. Also describes rightly the hierarchy of $V_{n}$ for small systems \cite{br1,br2,br3,br4,br5}.

\subsection{Second Phase Transition in the CSPM}
As the area covered by the strings 
 is $[1-exp(-\xi)]S_{\bot}$, the mean distance between strings $d$ is given by
\begin{equation}
  d = \left [\frac{N}{[1-exp(-\xi)]S_{\bot}} \right]^{-1/2}= F(\xi)\sqrt{\pi}r_{0}.
\end{equation}
which at the critical percolation density $\xi_{c} $ = 1.2, d~= 1.34$ r_{0}$.
This means that the overlapping of strings is very peripheral at this density,
covering only the edges(corona) of the strings.
 At this density using Eq.~(\ref{temp}), T is $\sim$ 160 MeV corresponding to the hadronization temperature and a partial deconfinement because the string
 structure is not destroyed, even though the color field is extended over
 almost  the whole collision surface.
 From the point of view of statistical mechanics the formed medium can form a fluid (such as it has been observed) only if the color sources, in our case
 strings, have a hard core. If not, the radial distribution of the color sources either obeys the periodic structure of a solid or the flat distribution corresponding to a free gas \cite{ramirez}. 
 The size of this hard core $h$ provide us with a new scale. In order to penetrate the hard core the overlapping strings should be larger, in such a way that
 $d=2h < r{_0}$. For reasonable values of $h \sim  0.4-0.5 r_{0}$
 (half of the radius of the string), we have $F(\xi) \sim 0.45-0.50$ corresponding to the temperature of 210-220 MeV according to Eq.~(\ref{temp}).
  It is worth to ask for the string density needed to obtain percolation of the hard core of the strings.
  The corresponding $\xi$  will be larger than the critical string percolation
  $\xi{_c}$=1.2 by the factor ${(r_{0}/h)}^{2}$ 
  which for reasonable values of $h$ around $r_{0}/2$ gives 4.
  As the temperature is proportional to $\xi^{1/4}$ we have T=1.414$T_{c}$ as
  $T_{c}$ is around 155 MeV T is 220 MeV, similar value of the temperature obtained before, as it should be.
  The repulsion of strings have been studied previously in phenomenological models \cite{altsy,bierl1}. In the Lund-Pythia model is introduced as
  $``string shoving''$ which improve the agreement with some data concerning the elliptic flow and heavy flavor production in $pp$ and A-A collisions
  \cite{bierl1}. In the case of high string densities, each string have in its
  neighbor strings in all azimuthal directions in such a way that the net effect
  of the repulsion would be zero or negligible. However at low multiplicities
  in $pp$ and  in peripheral A-A collisions at low $p_{t}$ the repulsion can
  produce negative elliptic flow $v_{2}$ contrary to hydrodinamical models \cite{bierl2}.

  The value of $T\sim$ 220 MeV for the temperature can be taken with caution due to the uncertainties, mainly coming from the value of the size of the hard
  core, taken as half of the string transverse size. Some evaluations of
  penetration depth would favor a smaller core giving rise to higher T. Most of the proposals mentioned before obtain T= 285-300 MeV. It is well known that
  quenched QCD has a first order phase transition at T= 285 MeV. However the value of T$\sim$ 220 MeV is not only obtained as percolation of hard cores but also corresponds to the sudden rise of the $\varepsilon/T^{4}$ of Figs. 1 and 3.

  We have been concerned with the CSPM but most of the string models should
  incorporate some of the effects discussed here. In particular the glasma
  picture of the Color Glass Condensate (CGC) that can be related to the CSPM.
  In particular the size of the color flux tube is proportional to $1/Q_{s}$,
  determining the particle correlations. The color flux tube size increases as the energy or centrality increases, becoming smaller without any bound. If flux tubes of glasma have hard core , the size of the flux tube can not be smaller than the size of the hard core.

  Above T= 220 MeV we expect that some of the signatures of the fluid found above T = 160 MeV to disappear. In particular the elliptic flow and rest of the harmonics or long range rapidity correlations should diminish, although some other effects not related with collective motion can give contribution to those
  observable.
  The fact this new phase transition is not supported by the LQCD, can be due
  to a insufficient large volume as it is mentioned in the reference \cite{alex4}
  or to the lack of an exact confinement order parameter in spite of several
  unsuccessful theoretical attempts \cite{detar,weiss,faber,ghanbar}.


\begin{figure}[thbp]
\centering        
\vspace*{1.0cm}
\includegraphics[width=0.60\textwidth,height=3.5in]{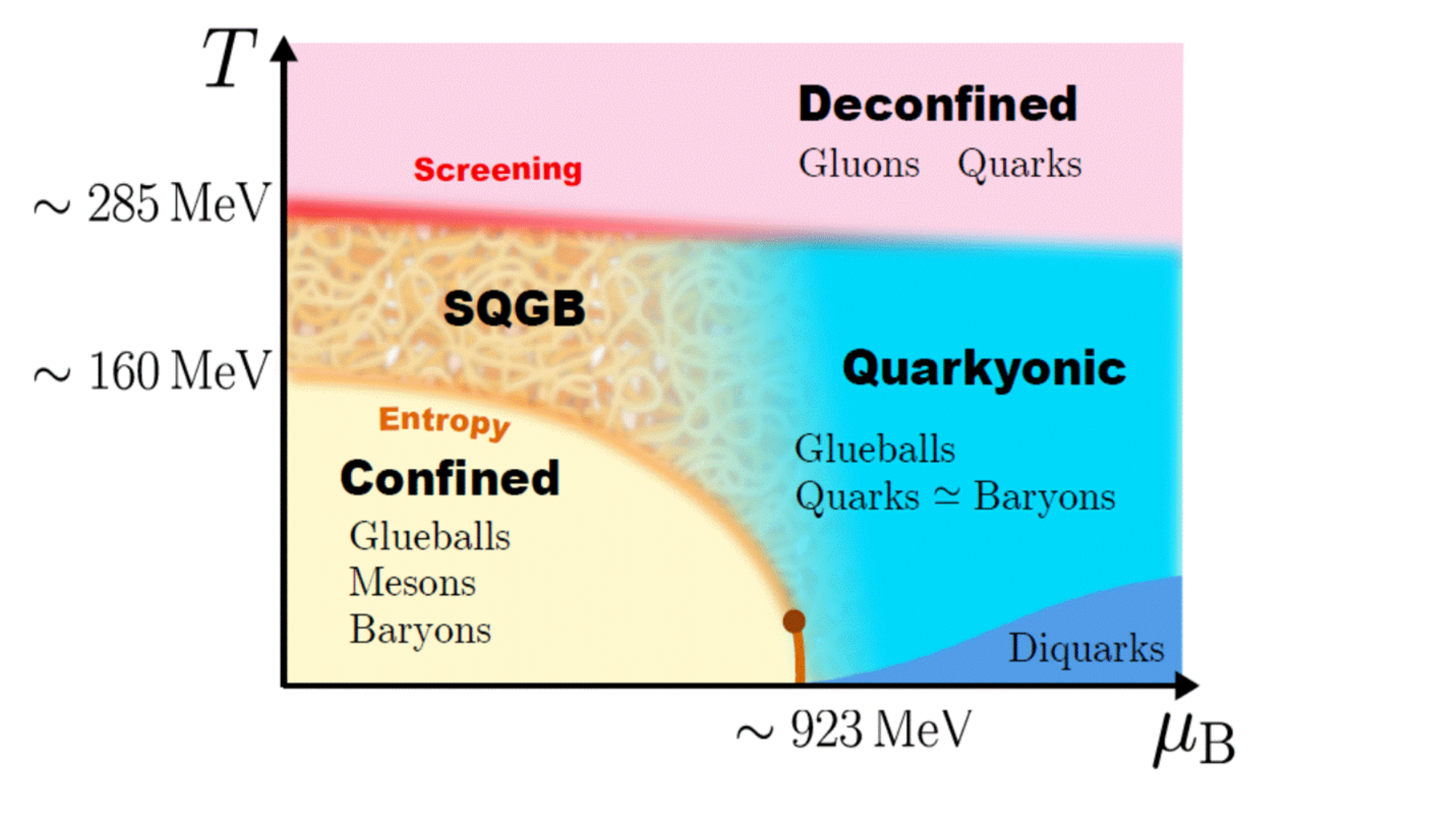}
\vspace*{-0.2cm}
\caption{New and realistic phase diagram for $N_{c}$ =3 including a window of the SQGB regime \cite{sqgp1}.}
\label{sqgb}
\end{figure}
\section{Conclusions}
  We explore the possibility to accommodate in the CSPM the new phase transition proposed in theoretical approaches. As the strings are color conductor repel the color field of other strings, meaning that they have a hard core. This
  structure is also required for the radial distribution of the color string sources in the transverse plane of the collision. In the fluid phase, this distribution can not be periodic (case of a solid) neither flat (case of a gas) but somthing like a hard core. A second critical density arises when the hard core of the strings cross the transverse surface of the collision, which means the percolation of the hard core of the strings. This critical string density corresponds a critical temperature of T $\sim$ 220 MeV, which marks the beginning of the transition from a stringy fluid to a free gas of deconfined quarks and gluons. This temperature coincides with the value of the sudden rise of $\varepsilon/T^{4}$
  evaluated from the ALICE data \cite{aditya} and also with the sudden rise of a string based model \cite{sqgp2}.   
\begin{figure}[!h]
\centering        
\includegraphics[width=0.55\textwidth,height=3.5in]{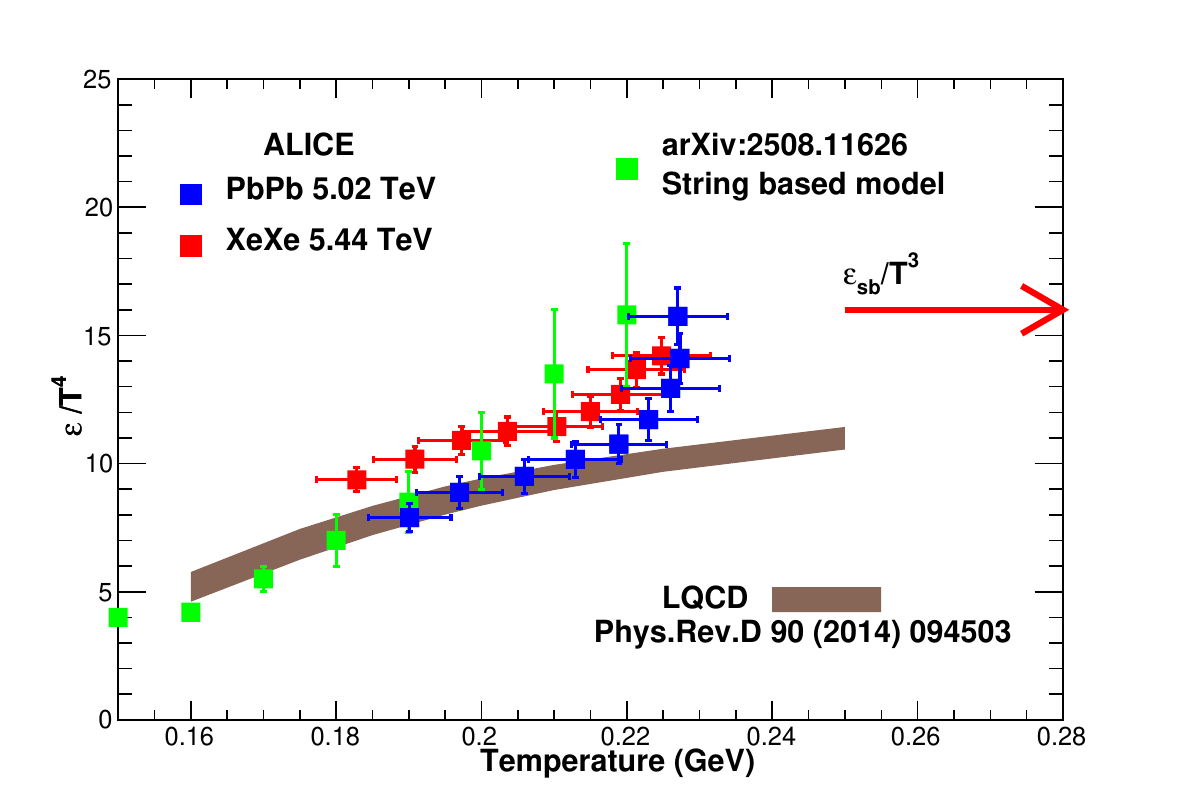}
 \vspace{-0.2cm}
 \caption{ Dimensionless quantity $\bm \varepsilon/T^{4}$ as a function of temperature from CSPM (using ALICE data) are shown as blue and red boxes \cite{aditya}. LQCD calculation from HotQCD Collaboration are shown as brown band \cite{lattice14} while String model results as green box \cite{sqgp2}.}
\label{sqgp}
\end{figure}
%

 \section{Acknowledgments}
 C. P. thanks the grant Maria de Maeztu excellence center of Spanish Science Ministry and ERDF of the European Union and by the Spanish Research Agency, Centro Singular de Galicia, 2023-2027 of Xunta de Galicia. We thank L. McLerran and
 N. Armesto  for stimulating discussions.   

\end{document}